\documentclass[11pt,titlepage,preprint]{article}
\usepackage{setspace}
\usepackage{amsmath}
\usepackage[dvips]{graphicx,epsfig}
\usepackage{jtb}

\begin{document}

\title{\bf Beneficial effects of intercellular interactions between pancreatic islet cells in blood glucose regulation}

\author{Junghyo Jo \\
    Laboratory of Biological Modeling, \\
    National Institute of Diabetes and Digestive and Kidney Diseases,\\	
    National Institutes of Health, Bethesda, MD 20892, U.S.A.\\
\and Moo Young Choi \thanks{
           Corresponding author.  Address:
           Department of Physics and Astronomy,
           Seoul National University, Seoul 151-747, Korea.
           E-mail: mychoi@snu.ac.kr }\\
    Department of Physics and Astronomy and Center for Theoretical Physics,\\
    Seoul National University, Seoul 151-747, Korea\\
\and Duk-Su Koh \\
    Department of Physiology and Biophysics,\\
    University of Washington, Seattle, WA 98195, U.S.A.
}


\maketitle

\abstract{
Glucose homeostasis is controlled by the islets of Langerhans
which are equipped with $\alpha$-cells increasing the blood glucose
level, $\beta$-cells decreasing it, and $\delta$-cells the precise
role of which still needs identifying.  Although intercellular
communications between these endocrine cells have recently been
observed, their roles in glucose homeostasis have not been clearly
understood.
In this study, we construct a mathematical model for an islet
consisting of two-state $\alpha$-, $\beta$-, and $\delta$-cells, and
analyze effects of known chemical interactions between them
with emphasis on the combined effects of those interactions.
In particular, such features as paracrine signals of neighboring cells and
cell-to-cell variations in response to external glucose concentrations
as well as glucose dynamics, depending on insulin and glucagon hormone, 
are considered explicitly.
Our model predicts three possible benefits
of the cell-to-cell interactions: First, the asymmetric
interaction between $\alpha$- and $\beta$-cells contributes to the
dynamic stability while the perturbed glucose level recovers to
the normal level.
Second, the inhibitory interactions of
$\delta$-cells for glucagon and insulin secretion prevent the
wasteful co-secretion of them at the normal glucose level.
Finally, the glucose dose-responses of insulin secretion is modified
to become more pronounced at high glucose levels due to the inhibition
by $\delta$-cells. 
It is thus concluded that the intercellular communications in islets of Langerhans
should contribute to the effective  
control of glucose homeostasis.

}

\clearpage

\section{Introduction}
Homeostasis, maintenance of the constant physiological state, is one
of the main characteristics of life. In particular, glucose
homeostasis is critical because glucose is the energy source for our
bodies; the malfunctioning of this process causes several disease
states including diabetes mellitus and brain coma.

In order to understand glucose homeostasis, we first need to examine
the tissue controlling the blood glucose level, the islet of
Langerhans in the pancreas.  It consists mainly of three types of
endocrine cells: $\alpha$-cells which secrete glucagon hormone
increasing the glucose level, $\beta$-cells which secrete insulin
decreasing the glucose level, and $\delta$-cells which secrete somatostatin,
known to inhibit activities of $\alpha$- and $\beta$-cells. The
hormone secretion of a cell influences the behavior of neighboring
cells, and is thus tightly correlated with the islet
structure~\cite{Hopcroft,Pipeleers2}. In rodents, an islet contains
about 1,000 endocrine cells on average: $\beta$-cells, occupying the
most volume (70 to 80\%) of an islet, populate largely in its core,
whereas non-$\beta$-cells are located on the mantle~\cite{Brissova}.

To the first approximation, $\alpha$- and $\beta$-cells should be
sufficient for glucose control because $\alpha$-cells can increase
the glucose level whereas $\beta$-cells can decrease the level. The
importance of this bi-hormonal mechanism for glucose homeostasis has
been well recognized~\cite{Cherrington}. However, it should be noted
that endocrine cells in the islet interact with each other rather than
act independently.  For example, the electrical coupling
between $\beta$-cells through gap-junctions is known to enhance
insulin secretion of coupled $\beta$-cells~\cite{Jo,Pipeleers2,Sherman1}.
In addition, it has been recently reported that chemical interactions
between neighboring cells through
hormones~\cite{Cherrington,Franklin2,Orci,Ravier,Samols1,Samols2,Soria}
and neurotransmitters~\cite{Brice,Franklin1,Gilon,Moriyama,Rorsman,Wendt},
termed ``paracrine interaction,'' affect glucose regulation.

Among these intercellular communications, enhancement of insulin
secretion by glucagon~\cite{Brereton,Samols1,Soria} seems to be
paradoxical because $\alpha$-cells, playing the reciprocal role to
$\beta$-cells in glucose regulation, promote the activity of $\beta$-cells.
In contrast, insulin, secreted by $\beta$-cells, inhibits glucagon secretion
of $\alpha$-cells~\cite{Cherrington,Franklin2,Ravier,Samols2,Soria},
which appears natural. Furthermore, the role of the third cell-type,
$\delta$-cells, is still not completely known although there have
been reports that somatostatin hormone, secreted by $\delta$-cells,
suppresses the hormone secretion of both $\alpha$- and
$\beta$-cells~\cite{Cherrington,Daunt,Orci,Soria}.

What is then the raison d'etre of the paradoxical interactions
between $\alpha$- and $\beta$-cells and the inhibitory action 
by $\delta$-cells?  Despite previous 
studies as to these questions over
the last thirty years~\cite{Orci,Pipeleers3,Soria,Unger}, there
still lacks concrete understanding of the role of these interactions
in terms of glucose homeostasis. The primary difficulty in understanding
these interactions lies in the complexity of the islet system which
includes many interactions between different coexisting cell-types
working in different conditions.

In this paper, we analyze the interactions between $\alpha$-,
$\beta$-, and $\delta$-cells, which contribute to the precise
control of the glucose level, by means of a mathematical model
incorporating experimentally known interactions between islet cells
(see above). As a result, our model predicts that the
intracellular interactions modify insulin and glucagon secretion
in a way to control the blood glucose level more efficiently.


\section{Islet model}
\subsection{Activity of islet cells}
We begin with a simplified model in which cells of each type can
take one of two states (active and silent). The state of a cell,
represented by ``Ising spin'' $\sigma$, is defined to be active
($\sigma=+1$) when the cell secretes islet hormone; otherwise the
state is defined as silent ($\sigma=-1$). 
Accordingly, the state of an islet consisting of $\alpha$-,
$\beta$-, and $\delta$-cells can be represented by
($\sigma_{\alpha}, \sigma_{\beta}, \sigma_{\delta}$). There is a
total of $2^3$ possible states of the islet, among which ($+1, -1,
-1$) and ($-1, +1, +1$) describe the islet state at low and high
glucose levels, respectively. The main source of changing cell states
is the blood glucose level $\tilde{G}$, which globally influences
all cells. In addition, the paracrine interaction $\tilde{J}$ from
neighboring cells locally affects cell states. In this manner we
obtain a simple Ising-type model, generally characterizing two-state
dynamics in statistical physics: The glucose level $\tilde{G}$
corresponds to the external magnetic field and the paracrine
interaction $\tilde{J}$ to the local interaction between spins.

\begin{figure}[h]
\centerline{
\includegraphics*[width=0.5\textwidth]{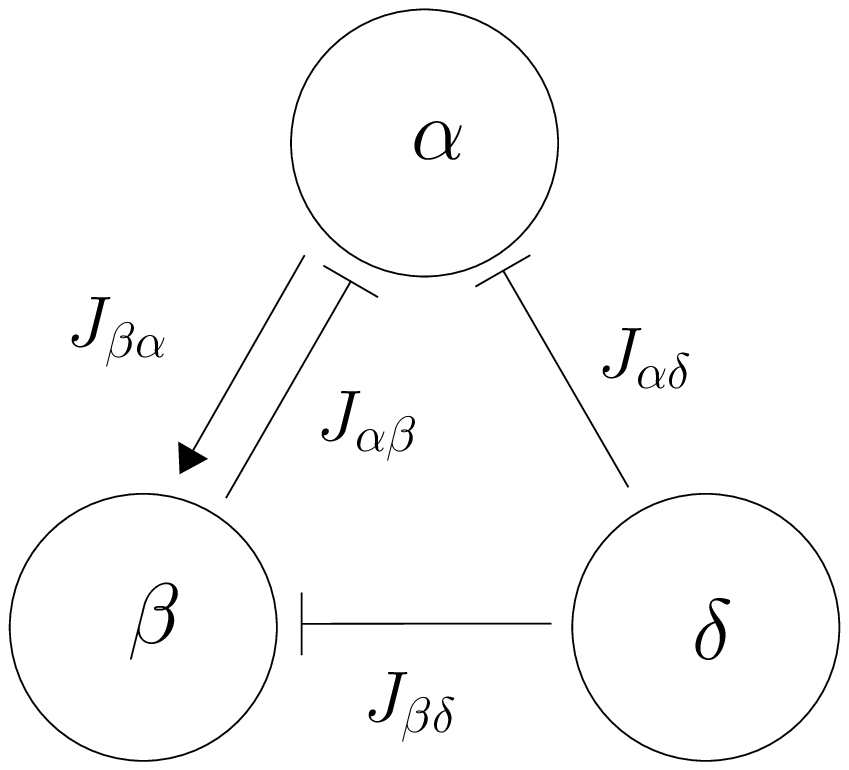}
}
\caption[]{}
\label{fig:interaction_abd}
\end{figure}

>From the known cellular interactions illustrated in
Fig.~\ref{fig:interaction_abd}, one may determine local stimuli
$G_{\alpha}$, $G_{\beta}$, and $G_{\delta}$, which change the states
of $\alpha$-, $\beta$-, and $\delta$-cells, respectively, in the
forms:
\begin{eqnarray} \label{eq:2localf_a}
G_{\alpha}&=&- G - \frac{1+\sigma_{\beta}}{2} J_{\alpha\beta} - \frac{1+\sigma_{\delta}}{2} J_{\alpha\delta} \nonumber \\
G_{\beta}&=&G+ \frac{1+\sigma_{\alpha}}{2} J_{\beta\alpha} -
\frac{1+\sigma_{\delta}}{2} J_{\beta\delta}  \nonumber \\ 
G_{\delta}&=&m G,
\end{eqnarray}
where $G \equiv \tilde{G}- \tilde{G}_0$ measures the excess glucose level
from the basal glucose level $\tilde{G}_0$ during the fasting period. 
The reciprocal nature of $\alpha$- and $\beta$-cells in the
responses to glucose is manifested by the opposite signs in front of $G$
in the first equation (for $G_{\alpha}$) and the second one (for
$G_{\beta}$) of Eq.~\ref{eq:2localf_a}. 
In addition, the asymmetric interaction between these two cell types
is also reflected 
in the second terms involving $J_{\alpha\beta}$ and $J_{\beta\alpha}$ of the equations. 
For simplicity, we assume that the interaction strength $J_{\beta\alpha}$ from $\alpha$- to $\beta$-cells
is the same as $J_{\alpha\beta}$ from $\beta$- to $\alpha$-cells and given by $J_1$, 
i.e., $J_{\alpha\beta}=J_{\beta\alpha}=J_1$. 
The last terms involving $J_{\alpha\delta}$ and $J_{\beta\delta}$ describe the inhibition effects 
of $\delta$-cells on $\alpha$- and $\beta$-cells, both with negative signs. 
Although the endogenous strengths of the interactions from $\delta$-cells to $\alpha$- and $\beta$-cells
are not known, the exogenous stimulus of somatostatin has been reported to inhibit both insulin 
and glucagon secretion to a similar degree~\cite{Cherrington}.
As a first approximation, it is thus assumed that both interactions have the same strength: 
$J_{\alpha\delta} =J_{\beta\delta}=J_2$. 
Here the interaction strengths $J_1$ and $J_2$ are expressed in terms of the relative effects to glucose
stimulation, and therefore have the unit of mM corresponding to the hormonal stimulus $\tilde{J}$, namely,
a given amount of stimulus $\tilde{J}$ by hormone is considered to produce the same
effects on a cell as a certain amount $J$ of glucose stimulation. 
In our simplified model, $\delta$-cells are not influenced by neighboring
$\alpha$- and $\beta$-cells but stimulated solely by glucose; therefore,
$G_{\delta}$ depends only on $G$ in Eq.~\ref{eq:2localf_a}. Like
$\beta$-cells, $\delta$-cells become active,
and secrete somatostatin above a threshold level of
glucose. The glucose sensitivity of $\delta$-cells is expected to
have a value between zero and unity, i.e., $0<m<1$ because the threshold level for the
activation of $\delta$-cells is lower than that of
$\beta$-cells~\cite{Efendic,Nadal}. We thus choose
the value $m=0.5$
in this study; the overall behavior does not depend qualitatively
on the value of $m$.

For given local stimulus $G_\alpha$ considering glucose stimulus $G$ and effects of insulin 
and somatostatin, the transition rate of an $\alpha$-cell
from state $\sigma_{\alpha}$ to state $-\sigma_{\alpha}$ depends on the states of other cell types 
as well as its own state, and is denoted as $w_{\alpha}(\sigma_{\alpha}, \sigma_{\beta}, \sigma_{\delta})$.
This transition rate should satisfy the detailed balance condition between two $\alpha$-cell states 
$\sigma_\alpha$ and $-\sigma_\alpha$ at equilibrium:
\begin{equation}
\label{eq:detailed_balance}
w_\alpha (\sigma_\alpha, \sigma_\beta, \sigma_\delta) P(\sigma_\alpha)=
w_\alpha (-\sigma_\alpha, \sigma_\beta, \sigma_\delta) P(-\sigma_\alpha),
\end{equation}
where the probability $P(\sigma_\alpha)$ for state $\sigma_\alpha$ follows the Boltzmann distribution
$\exp [- G_\alpha (1+\sigma_\alpha)/2\Theta]$ with respect to the quantity
$G_\alpha (1+\sigma_\alpha)/2$ for the local stimulus $G_\alpha$. 
Namely, $\alpha$-cells favor the state minimizing the quantity $G_\alpha (1+\sigma_\alpha)/2$. 
Here $\Theta$ measures the amount of uncertainty, which is inevitable in biological
systems.  The origin may be the heterogeneous glucose sensitivity of cells 
and/or the environmental noise including thermal fluctuations.  It is obvious that 
$G_\alpha (1+\sigma_\alpha)/2$ and $\Theta$ correspond to the energy and the temperature, 
respectively, in statistical physics. 
In this study, the ``temperature'' is taken to be unity ($\Theta = 1$) in units of the ``energy'', 
which is biologically tantamount to the fluctuations caused by 1\,mM change of glucose stimulation.  

The ratio between the reciprocal transition rates thus reads
\begin{eqnarray}
\label{eq:2tr} \frac{w_{\alpha}(\sigma_{\alpha}, \sigma_{\beta},
\sigma_{\delta})} {w_{\alpha}(-\sigma_{\alpha}, \sigma_{\beta},
\sigma_{\delta})}& = & \exp \Bigg[-\frac{1}{\Theta}
G_{\alpha} \sigma_\alpha \Bigg] \nonumber \\
& = & \exp \Bigg[\frac{1}{\Theta} \Bigg(G_\alpha^{\text {eff}} \sigma_\alpha
+ \frac{J_1}{2} \sigma_\alpha \sigma_\beta
+ \frac{J_2}{2} \sigma_\delta \sigma_\alpha \Bigg) \Bigg] 
\end{eqnarray}
with $G_\alpha^{\text {eff}} \equiv G + J_1/2 + J_2/2$, 
where Eq.~\ref{eq:2localf_a} has been used to obtain the second line. 
There the three stimulation terms represent effective glucose stimulation, 
paracrine interaction from $\beta$-cells, and another from $\delta$-cells, respectively. 
Assuming that these stimuli affect independently the $\alpha$-cell state, 
we write the transition rate in the form
\begin{eqnarray}
\label{eq:glauber} w_{\alpha} 
&=& \frac{1}{2 \tau} \left[1+\tanh\left(\frac{G_\alpha^{\text {eff}}}{2 \Theta} \right) \sigma_{\alpha} \right]
\left[1+\tanh\left(\frac{J_1}{4 \Theta} \right) \sigma_{\alpha}\sigma_{\beta} \right] \nonumber \\
&& \times \left[1+\tanh\left(\frac{J_2}{4 \Theta}
\right) \sigma_{\delta} \sigma_{\alpha} \right],
\end{eqnarray}
where $\tau$ measures the characteristic time of the transition and it has been noted that 
$\tanh(y \sigma) = \sigma \tanh y$ for $\sigma = \pm 1$. 
Note that among possible transition rates satisfying Eq.~\ref{eq:detailed_balance}, 
we adopt the Glauber dynamics~\cite{Glauber} to choose the specific form of Eq.~\ref{eq:glauber}, 
which exhibits the sigmoidal form ubiquitously describing response functions in biological systems. 
However, the behavior of the system in general does not depend qualitatively on the
specific form of the transition rate satisfying Eq.~\ref{eq:detailed_balance}. 
Similarly, we obtain the transition rates $w_\beta$ and $w_\delta$ of 
$\beta$- and $\delta$-cells. 

The transition rates of three cell types can be summarized as
\begin{eqnarray}
\label{eq:2tr2} w_x (\sigma_{\alpha}, \sigma_{\beta},
\sigma_{\delta})&=&\frac{1}{2 \tau} \left[w^x + w^x_{\alpha}
\sigma_{\alpha}
+ w^x_{\beta} \sigma_{\beta} + w^x_{\delta} \sigma_{\delta}
+ w^x_{\alpha \beta} \sigma_{\alpha} \sigma_{\beta}\right. \nonumber \\
&&\left. +w^x_{\beta \delta} \sigma_{\beta} \sigma_{\delta}
+w^x_{\delta \alpha} \sigma_{\delta} \sigma_{\alpha} +w^x_{\alpha
\beta \delta} \sigma_{\alpha} \sigma_{\beta} \sigma_{\delta}
\right],
\end{eqnarray}
with $x=\alpha, \beta,$ and $\delta$, where the coefficients are
given in Tables~\ref{tab:coeff} and \ref{tab:param}.
%
The master equation, describing the evolution of the probability
$P(\sigma_{\alpha}, \sigma_{\beta}, \sigma_{\delta})$ for the islet in state
$(\sigma_{\alpha}, \sigma_{\beta}, \sigma_{\delta})$, reads
\begin{eqnarray}
\label{eq:master_abd} && \frac{d}{dt}P(\sigma_{\alpha},
\sigma_{\beta}, \sigma_{\delta}) \nonumber \\
&& ~~= w_{\alpha}(-\sigma_{\alpha}, \sigma_{\beta}, \sigma_{\delta})
P(-\sigma_{\alpha}, \sigma_{\beta}, \sigma_{\delta})
 +w_{\beta}(\sigma_{\alpha}, -\sigma_{\beta}, \sigma_{\delta})
P(\sigma_{\alpha}, -\sigma_{\beta}, \sigma_{\delta})
\nonumber \\
 && ~~~~~+w_{\delta}(\sigma_{\alpha}, \sigma_{\beta}, -\sigma_{\delta})
P(\sigma_{\alpha}, \sigma_{\beta}, -\sigma_{\delta})
-w_{\alpha}(\sigma_{\alpha}, \sigma_{\beta}, \sigma_{\delta})
P(\sigma_{\alpha}, \sigma_{\beta}, \sigma_{\delta})
\nonumber \\
 && ~~~~~-w_{\beta}(\sigma_{\alpha}, \sigma_{\beta}, \sigma_{\delta})
P(\sigma_{\alpha}, \sigma_{\beta}, \sigma_{\delta})
 - w_{\delta}(\sigma_{\alpha}, \sigma_{\beta}, \sigma_{\delta})
P(\sigma_{\alpha}, \sigma_{\beta}, \sigma_{\delta})
\end{eqnarray}
with the transition rates $w_{\alpha}$, $w_{\beta}$, and $w_{\delta}$ in Eq.~\ref{eq:2tr2}.
%
Note that Eq.~\ref{eq:master_abd} describes the net flux to state
$(\sigma_{\alpha}, \sigma_{\beta}, \sigma_{\delta})$ simply given by the difference
between the in-flux to state $(\sigma_{\alpha}, \sigma_{\beta}, \sigma_{\delta})$
from other states $(-\sigma_{\alpha}, \sigma_{\beta}, \sigma_{\delta})$,
$(\sigma_{\alpha}, - \sigma_{\beta}, \sigma_{\delta})$, and
$(\sigma_{\alpha}, \sigma_{\beta},  -\sigma_{\delta})$ and the out-flux
from state $(\sigma_{\alpha}, \sigma_{\beta}, \sigma_{\delta})$ to others.

>From this master equation, it is straightforward to obtain the time evolution of the ensemble
averages of the cell states and their correlations.
For example, multiplying both sides of Eq.~\ref{eq:master_abd} by $\sigma_{\alpha}$ and
summing over all configurations, we obtain the evolution equation for the average
$\langle \sigma_{\alpha} \rangle \equiv \sum_{\sigma_{\alpha}, \sigma_{\beta}, \sigma_{\delta}}
\sigma_{\alpha} P(\sigma_{\alpha}, \sigma_{\beta}, \sigma_{\delta})$
of the state of $\alpha$-cells:
\begin{equation}
\label{eq:eom_a}
\frac{d}{dt} \langle \sigma_{\alpha} \rangle =
 -2 \langle \sigma_{\alpha} w_{\alpha}(\sigma_{\alpha}, \sigma_{\beta}, \sigma_{\delta} )\rangle
\end{equation}
and similarly,
\begin{eqnarray}
\label{eq:eom_b_d}
\frac{d}{dt} \langle \sigma_{\beta} \rangle &=&
 -2 \langle \sigma_{\beta} w_{\beta}(\sigma_{\alpha}, \sigma_{\beta}, \sigma_{\delta} )\rangle \\
\frac{d}{dt} \langle \sigma_{\delta} \rangle &=&
 -2 \langle \sigma_{\delta} w_{\alpha}(\sigma_{\alpha}, \sigma_{\beta}, \sigma_{\delta} )\rangle.
\end{eqnarray}
Note that $1+\langle\sigma_{\alpha}\rangle$ gives twice the average activity of $\alpha$-cells, etc.

Among the eight equations for the probability $P(\sigma_{\alpha},
\sigma_{\beta}, \sigma_{\delta} )$ corresponding to the eight
possible states of the islet, only seven are independent, due to the
normalization condition $\sum_{\sigma_{\alpha}, \sigma_{\beta},
\sigma_{\delta}} p(\sigma_{\alpha}, \sigma_{\beta}, \sigma_{\delta}
)=1$. Therefore, there exist four more equations in addition to the
above three describing the average of cell states.  Those are
evolution equations for correlations of two cell states and of three
cell states.
The equation for the correlation function $\langle \sigma_{\alpha}
\sigma_{\beta} \rangle$ of the $\alpha$-cell and $\beta$-cell states
can again be derived from Eq. \ref{eq:master_abd}, multiplied by
$\sigma_{\alpha} \sigma_{\beta}$ and summed over all configurations:
\begin{equation}
\label{eq:eom_ab} \frac{d}{dt} \langle \sigma_{\alpha}
\sigma_{\beta} \rangle = -2 \langle \sigma_{\alpha} \sigma_{\beta}
w_{\alpha}(\sigma_{\alpha}, \sigma_{\beta}, \sigma_{\delta}
)\rangle -2 \langle \sigma_{\alpha} \sigma_{\beta}
w_{\beta}(\sigma_{\alpha}, \sigma_{\beta}, \sigma_{\delta}
)\rangle.
\end{equation}
The equations for  $\langle \sigma_{\beta} \sigma_{\delta} \rangle$
and $\langle \sigma_{\delta} \sigma_{\alpha} \rangle$ are also
obtained in the same way:
\begin{eqnarray}
\label{eq:eom_bd_da} \frac{d}{dt} \langle \sigma_{\beta}
\sigma_{\delta} \rangle &=&
 -2 \langle \sigma_{\beta} \sigma_{\delta} w_{\beta}(\sigma_{\alpha}, \sigma_{\beta}, \sigma_{\delta} )\rangle
 -2 \langle \sigma_{\beta} \sigma_{\delta} w_{\delta}(\sigma_{\alpha}, \sigma_{\beta}, \sigma_{\delta} )\rangle \nonumber\\
\frac{d}{dt} \langle \sigma_{\delta} \sigma_{\alpha} \rangle &=&
 -2 \langle \sigma_{\delta} \sigma_{\alpha} w_{\delta}(\sigma_{\alpha}, \sigma_{\beta}, \sigma_{\delta} )\rangle
 -2 \langle \sigma_{\delta} \sigma_{\alpha} w_{\alpha}(\sigma_{\alpha}, \sigma_{\beta}, \sigma_{\delta} )\rangle.
\end{eqnarray}
Note that correlations of two cell states represent the relative
activity of the two cells. Accordingly, it makes a good measure of
the different responses between two cells. Similarly, the equation
for correlations of three cell states is given by
\begin{eqnarray}
\label{eq:eom_abd} \frac{d}{dt} \langle \sigma_{\alpha}
\sigma_{\beta} \sigma_{\delta} \rangle &=&
 -2 \langle \sigma_{\alpha} \sigma_{\beta} \sigma_{\delta} w_{\alpha}(\sigma_{\alpha}, \sigma_{\beta}, \sigma_{\delta} )\rangle
 -2 \langle \sigma_{\alpha} \sigma_{\beta} \sigma_{\delta} w_{\beta}(\sigma_{\alpha}, \sigma_{\beta}, \sigma_{\delta} )\rangle \nonumber \\
&& -2 \langle \sigma_{\alpha} \sigma_{\beta} \sigma_{\delta}
w_{\delta}(\sigma_{\alpha}, \sigma_{\beta}, \sigma_{\delta}
)\rangle.
\end{eqnarray}

Substituting the transition rates in Eq.~\ref{eq:2tr2} into Eqs.
\ref{eq:eom_a} to \ref{eq:eom_abd}, we finally obtain equations for
the states of the three cell types and their correlations:
\begin{eqnarray}
\tau \frac{d}{dt} \langle \sigma_{\alpha} \rangle &=&
-w^{\alpha}_{\alpha} - w^{\alpha} \langle \sigma_{\alpha} \rangle
- w^{\alpha}_{\alpha \beta} \langle \sigma_{\beta} \rangle -
w^{\alpha}_{\delta \alpha} \langle \sigma_{\delta} \rangle
- w^{\alpha}_{\beta} \langle \sigma_{\alpha} \sigma_{\beta} \rangle \nonumber \\
&& - w^{\alpha}_{\alpha \beta \delta} \langle \sigma_{\beta}
\sigma_{\delta} \rangle
- w^{\alpha}_{\delta} \langle \sigma_{\delta} \sigma_{\alpha} \rangle + w^{\alpha}_{\beta \delta} \langle \sigma_{\alpha} \sigma_{\beta} \sigma_{\delta} \rangle \nonumber \\
\tau \frac{d}{dt} \langle \sigma_{\beta} \rangle &=&
-w^{\beta}_{\beta} - w^{\beta}_{\alpha \beta} \langle
\sigma_{\alpha} \rangle - w^{\beta} \langle \sigma_{\beta} \rangle
- w^{\beta}_{\beta \delta} \langle \sigma_{\delta} \rangle
- w^{\beta}_{\alpha} \langle \sigma_{\alpha} \sigma_{\beta} \rangle \nonumber \\
&& - w^{\beta}_{\delta} \langle \sigma_{\beta} \sigma_{\delta}
\rangle - w^{\beta}_{\alpha \beta \delta} \langle \sigma_{\delta}
\sigma_{\alpha} \rangle
- w^{\beta}_{\delta \alpha} \langle \sigma_{\alpha} \sigma_{\beta} \sigma_{\delta} \rangle  \nonumber \\
\tau \frac{d}{dt} \langle \sigma_{\delta} \rangle &=&
-w^{\delta}_{\delta} - w^{\delta}_{\delta \alpha} \langle
\sigma_{\alpha} \rangle - w^{\delta}_{\beta \delta} \langle
\sigma_{\beta} \rangle - w^{\delta}  \langle \sigma_{\delta}
\rangle
- w^{\delta}_{\alpha \beta \delta} \langle \sigma_{\alpha} \sigma_{\beta} \rangle \nonumber \\
&& - w^{\delta}_{\beta} \langle \sigma_{\beta} \sigma_{\delta}
\rangle - w^{\delta}_{\alpha} \langle \sigma_{\delta}
\sigma_{\alpha} \rangle
- w^{\delta}_{\alpha \beta} \langle \sigma_{\alpha} \sigma_{\beta} \sigma_{\delta} \rangle  \nonumber \\
\tau \frac{d}{dt} \langle \sigma_{\alpha} \sigma_{\beta} \rangle
&=& -(w^{\alpha}_{\alpha \beta}+w^{\beta}_{\alpha \beta})
-(w^{\alpha}_{\beta}+w^{\beta}_{\beta}) \langle \sigma_{\alpha}
\rangle
-(w^{\alpha}_{\alpha}+w^{\beta}_{\alpha}) \langle \sigma_{\beta} \rangle \nonumber \\
&& -(w^{\alpha}_{\alpha \beta \delta}+w^{\beta}_{\alpha \beta
\delta}) \langle \sigma_{\delta} \rangle
-(w^{\alpha}+w^{\beta}) \langle \sigma_{\alpha} \sigma_{\beta} \rangle \nonumber \\
&& - (w^{\alpha}_{\delta \alpha}+w^{\beta}_{\delta \alpha})
\langle \sigma_{\beta} \sigma_{\delta} \rangle
- (w^{\alpha}_{\beta \delta}+w^{\beta}_{\beta \delta}) \langle \sigma_{\delta} \sigma_{\alpha} \rangle \nonumber \\
&&- (w^{\alpha}_{\delta}+w^{\beta}_{\delta}) \langle \sigma_{\alpha} \sigma_{\beta} \sigma_{\delta} \rangle  \nonumber \\
\tau \frac{d}{dt} \langle \sigma_{\beta} \sigma_{\delta} \rangle
&=& -(w^{\beta}_{\beta \delta}+w^{\delta}_{\beta \delta}) -
(w^{\beta}_{\alpha \beta \delta}+w^{\delta}_{\alpha \beta \delta})
\langle \sigma_{\alpha} \rangle -
(w^{\beta}_{\delta}+w^{\delta}_{\delta}) \langle \sigma_{\beta}
\rangle \nonumber \\&& - (w^{\beta}_{\beta}+w^{\delta}_{\beta})
\langle \sigma_{\delta} \rangle
- (w^{\beta}_{\delta \alpha}+w^{\delta}_{\delta \alpha}) \langle \sigma_{\alpha} \sigma_{\beta} \rangle \nonumber \\
&&- (w^{\beta}+w^{\delta}) \langle \sigma_{\beta} \sigma_{\delta}
\rangle
- (w^{\beta}_{\alpha \beta}+w^{\delta}_{\alpha \beta}) \langle \sigma_{\delta} \sigma_{\alpha} \rangle \nonumber \\
&&- (w^{\beta}_{\alpha}+w^{\delta}_{\alpha}) \langle \sigma_{\alpha} \sigma_{\beta} \sigma_{\delta} \rangle  \nonumber \\
\tau \frac{d}{dt} \langle \sigma_{\delta} \sigma_{\alpha} \rangle
&=& -(w^{\delta}_{\delta \alpha}+w^{\alpha}_{\delta \alpha}) -
(w^{\delta}_{\delta}+w^{\alpha}_{\delta}) \langle \sigma_{\alpha}
\rangle
- (w^{\delta}_{\alpha \beta \delta}+w^{\alpha}_{\alpha \beta \delta}) \langle \sigma_{\beta} \rangle \nonumber \\
&&- (w^{\delta}_{\alpha}+w^{\alpha}_{\alpha}) \langle
\sigma_{\delta} \rangle
- (w^{\delta}_{\beta \delta}+w^{\alpha}_{\beta \delta}) \langle \sigma_{\alpha} \sigma_{\beta} \rangle \nonumber \\
&&- (w^{\delta}_{\alpha \beta}+w^{\alpha}_{\alpha \beta}) \langle
\sigma_{\beta} \sigma_{\delta} \rangle
- (w^{\delta}+w^{\alpha}) \langle \sigma_{\delta} \sigma_{\alpha} \rangle \nonumber \\
&&- (w^{\delta}_{\beta}+w^{\alpha}_{\beta}) \langle \sigma_{\alpha} \sigma_{\beta} \sigma_{\delta} \rangle  \nonumber \\
\tau \frac{d}{dt} \langle \sigma_{\alpha} \sigma_{\beta}
\sigma_{\delta}  \rangle &=& -(w^{\alpha}_{\alpha \beta
\delta}+w^{\beta}_{\alpha \beta \delta} +w^{\delta}_{\alpha \beta
\delta}) - (w^{\alpha}_{\beta \delta}+w^{\beta}_{\beta \delta}
+w^{\delta}_{\beta \delta}) \langle \sigma_{\alpha} \rangle
\nonumber \\
&& - (w^{\alpha}_{\delta \alpha}+w^{\beta}_{\delta \alpha}
+w^{\delta}_{\delta \alpha}) \langle \sigma_{\beta} \rangle -
(w^{\alpha}_{\alpha \beta}+w^{\beta}_{\alpha \beta}
+w^{\delta}_{\alpha \beta}) \langle \sigma_{\delta} \rangle
\nonumber \\
&& - (w^{\alpha}_{\delta}+w^{\beta}_{\delta} +w^{\delta}_{\delta})
\langle \sigma_{\alpha} \sigma_{\beta} \rangle
- (w^{\alpha}_{\alpha}+w^{\beta}_{\alpha} +w^{\delta}_{\alpha}) \langle \sigma_{\beta} \sigma_{\delta} \rangle \nonumber \\
&&- (w^{\alpha}_{\beta}+w^{\beta}_{\beta} +w^{\delta}_{\beta})
\langle \sigma_{\delta} \sigma_{\alpha} \rangle -
(w^{\alpha}+w^{\beta} +w^{\delta}) \langle \sigma_{\alpha}
\sigma_{\beta} \sigma_{\delta} \rangle . \label{eq:evol_abd}
\end{eqnarray}

\subsection{Glucose homeostasis}
Heretofore we have focused on the cellular interactions at a given
glucose level. To study dynamics of glucose homeostasis, however,
we should also take into account the change of the glucose level
and incorporate another equation for glucose regulation into the model.
Based on the fact that $\alpha$- and $\beta$-cells
secrete glucagon and insulin, respectively, raising and reducing
the glucose level, the equation for the glucose level $G$ is taken
to be
\begin{equation}
\label{eq:glucose} \tau_G \frac{dG}{dt}= \frac{1+\langle
\sigma_{\alpha} \rangle}{2}- \frac{1+\langle \sigma_{\beta}
\rangle}{2},
\end{equation}
where $\tau_G$ is the characteristic time for the hormones to
regulate the glucose level. It is expected that $\tau_G$ is larger
than the characteristic time $\tau$ of 
the change in cell states.
Equation \ref{eq:glucose} describes the decrease or increase of
the glucose level when $\alpha$-cells or $\beta$-cells are active
($\sigma_{\alpha}=1$ or $\sigma_{\beta}=1$).
Here, for simplicity, we have used the same characteristic time $\tau_G$
for glucagon and insulin to regulate glucose levels. 
Having different time constants turns out merely to shift the stationary level of blood glucose. 
%
To sum, we have a total of eight differential equations given by
Eqs.~\ref{eq:evol_abd} and \ref{eq:glucose}, which describe the
process of glucose homeostasis.

\section{Results}

\subsection{Asymmetric interactions between $\alpha$- and $\beta$-cells}
In our model, activities of $\alpha$-, $\beta$-, and
$\delta$-cells are determined by the external glucose level
together with feedback loops of intercellular 
interactions. A given
cell, subject to a glucose stimulus, secretes hormone which
influences the behavior of neighboring cells. In response, the
neighboring cells reversely influence the given cell.
These mutual interactions through hormones constitute the feedback
loop which is widely employed for advanced system control in
engineering~\cite{Bechhoefer}.

The interactions between $\alpha$- and $\beta$-cells are asymmetric:
While glucagon secreted from $\alpha$-cells enhances insulin
secretion of $\beta$-cells~\cite{Brereton,Samols1,Soria}, insulin
inhibits glucagon
secretion~\cite{Cherrington,Franklin2,Ravier,Samols2,Soria}. The
former positive interaction to the counterpart cells may seem
strange, but it eventually contributes to the construction of a
negative feedback loop for both cells. At low glucose levels,
$\alpha$-cells secrete glucagon, which enhances insulin secretion.
In turn, insulin inhibits the glucagon secretion of $\alpha$-cells.
Therefore, their interactions as a whole tend to suppress the glucagon
secretion from $\alpha$-cells. Similar negative feedback operates when $\beta$-cells are activated by high glucose concentration.
It is noteworthy that this feedback works more efficiently in case that the glucose level varies. 
At a static glucose level, it should be difficult for the mutual interactions between $\alpha$-
and $\beta$-cells to arise simultaneously because insulin and glucagon
are secreted at different glucose levels.

\begin{figure}[h]
\centerline{
        \includegraphics*[width=1\textwidth]{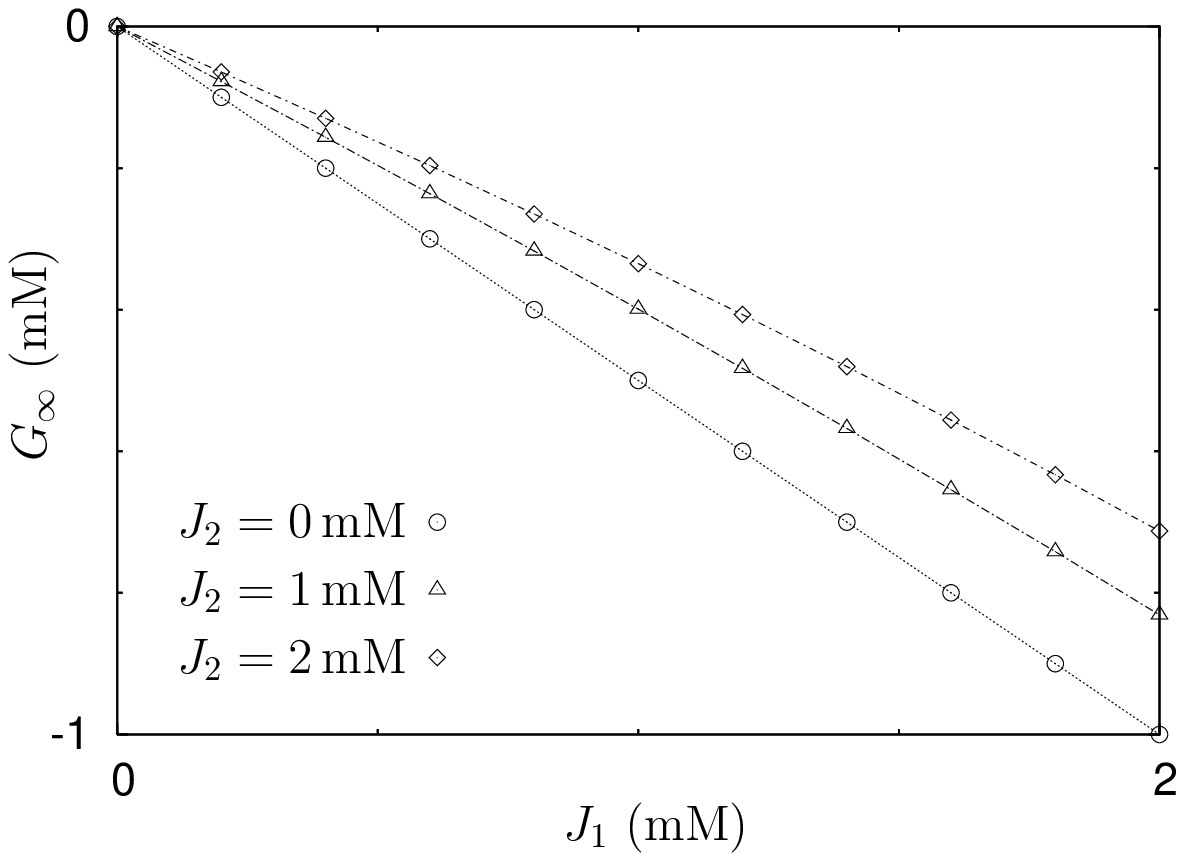}
}
        \caption[]{}
        \label{fig:stationary}
\end{figure}

In general, a negative feedback favors stability of a system because
it attenuates overaction of the system such as overshoot or undershoot.
The negative feedbacks in an islet system contribute to the stable
recovery to the normal glucose level $G_\infty$, when the system is
externally perturbed by stimuli such as a glucose dose. The normal
glucose level $G_\infty$, reached by $G \,(\equiv \tilde{G}-\tilde{G}_0 )$
at stationarity, depends on the cellular interactions shown in
Fig.~\ref{fig:stationary}. The asymmetric interaction $J_1$ lowers
the basal glucose level because $\alpha$-cells activate $\beta$-cells which
secret insulin and thus reduces the glucose level. In addition, the
inhibitory interaction $J_2$ of $\delta$-cells, albeit the same for $\alpha$- and $\beta$-cells, 
suppresses the activity of $\beta$-cells more than that of $\alpha$-cells at the
normal glucose level, 
because the activity of $\beta$-cells is higher than that of $\alpha$-cells resulting from the asymmetric interaction between $\alpha$- and $\beta$-cells.
Accordingly, the basal glucose level tends to increase 
as the strength $J_2$ of the inhibitory interaction is increased.

\begin{figure}[h]
\centerline{
        \includegraphics*[width=1\textwidth]{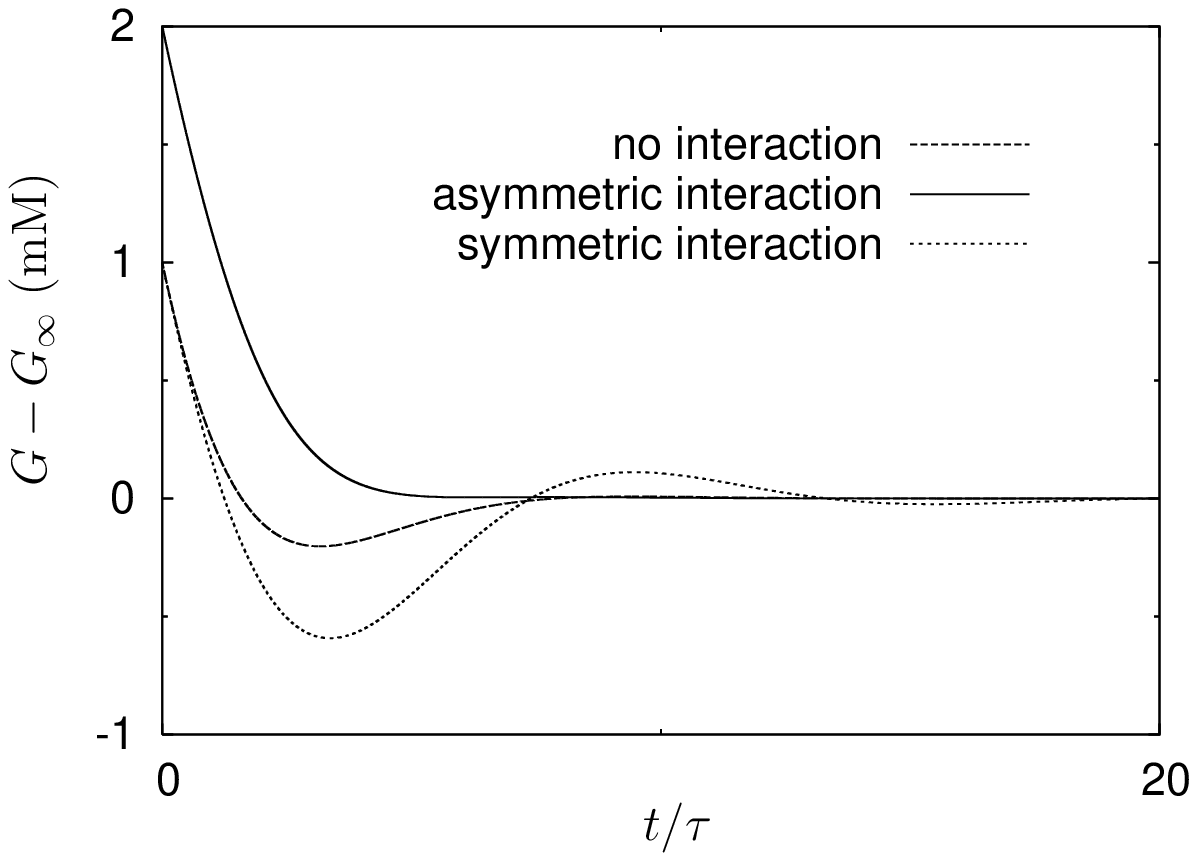}
}
        \caption[]{}
        \label{fig:glucose}
\end{figure}

Figure~\ref{fig:glucose} demonstrates the smooth recovery of the
glucose level in the presence of cellular interactions (solid
line), compared with the somewhat erratic recovery, once reaching
low glucose levels, in the absence of the interactions (dashed
line). For comparison, we also consider the behavior in the case of
symmetric interactions between $\alpha$- and $\beta$-cells, i.e.,
where glucagon inhibits insulin secretion and vice versa,
only to find even more erratic recovery (see the dotted line).  Shown
here is the recovery from the high glucose state [$G=1$\,mM (or
$\tilde{G} = \tilde{G}_0$ + 1\,mM), $\langle
\sigma_{\alpha} \rangle = -1$, and $\langle \sigma_{\beta} \rangle =1$].
The recovery from a low glucose state gives the same results
(data not shown) 
although such erratic recovery is more pronounced for the glucose level
starting from a higher value. 

\begin{figure}[h]
\centerline{
        \includegraphics*[width=1\textwidth]{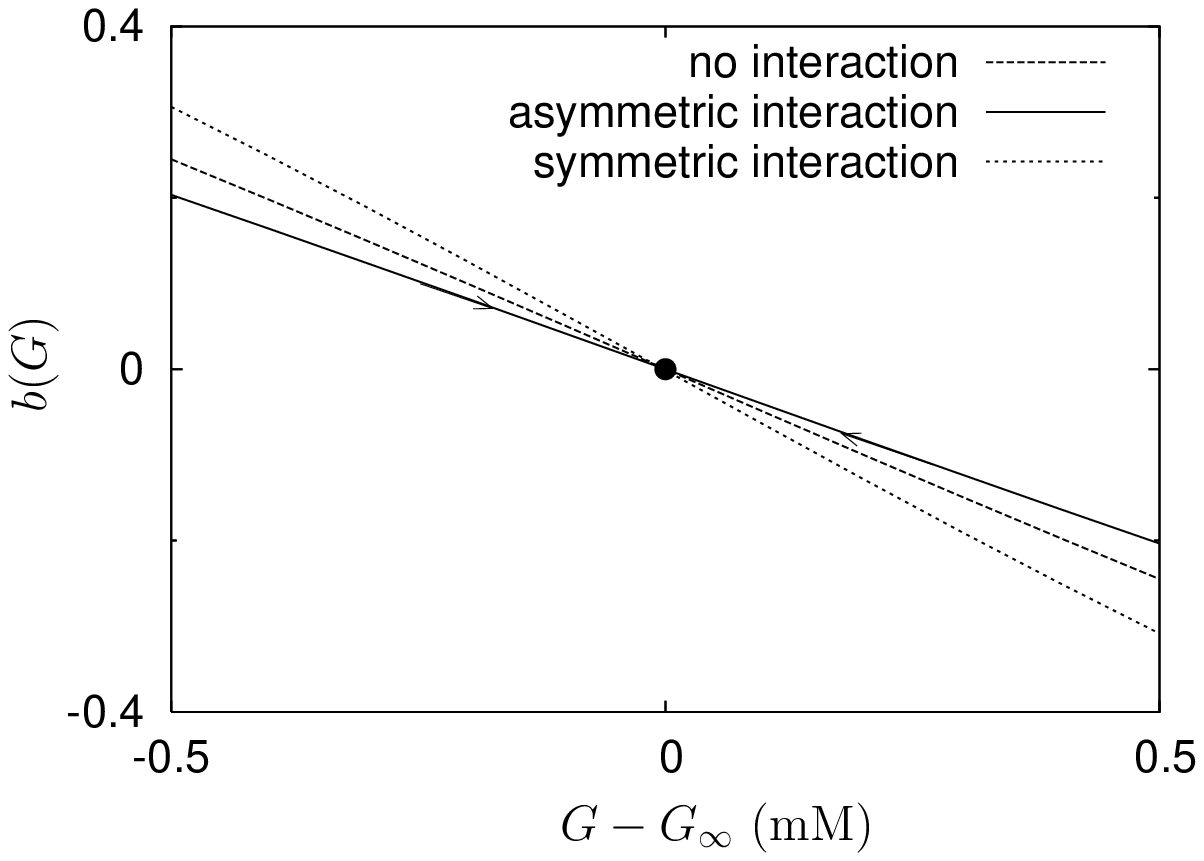}
}
        \caption[]{}
        \label{fig:balance}
\end{figure}

To examine the stability in approaching the normal glucose level,
we define the balance function
\begin{equation}
\label{eq:balance} b(G) \equiv \tau_G \frac{dG}{dt}=
\frac{1+\langle \sigma_{\alpha} \rangle}{2} - \frac{1+\langle
\sigma_{\beta} \rangle}{2},
\end{equation}
which describes the glucose level change during the characteristic
time. Since the activity of cells represents their hormone
secretion, $b(G)$ appropriately describes the effectiveness of the
glucose regulation by $\alpha$- and $\beta$-cells. If the
characteristic time $\tau_G$ of glucose regulation is much larger
than the characteristic time $\tau$ of cell responses in
Eq.~\ref{eq:evol_abd}, i.e., $\tau \ll \tau_G$, the glucose level
should be in a quasi-stationary state at time $t$ shorter than
$\tau_G$. Then the fast dynamics of cell states in
Eq.~\ref{eq:evol_abd} saturates rapidly at a given glucose level
and the seven variables, activities and correlations, reach their
fixed points depending on the glucose level $G$. In particular
$\langle \sigma_{\alpha} \rangle$ and $\langle \sigma_{\beta}
\rangle$ depend on $G$, giving the balance function in
Eq.~\ref{eq:balance} as a function of $G$, with a fixed point at
$G=G_\infty$ (see Fig.~\ref{fig:balance}). At low glucose levels
($G < G_\infty$), we have the balance function greater than zero
($b>0$), or $dG/dt > 0$, thus the glucose level grows with time.
At high glucose levels ($G > G_\infty$), the opposite behavior arises.
The resulting flow of the balance function is illustrated by the
arrows in Fig.~\ref{fig:balance} and it is concluded that the
balance function correctly describes glucose homeostasis. Further,
the slope of $b(G)$ near the fixed point $G=G_\infty$ represents
how smoothly the glucose level approaches the normal level: The
slope of the balance function for the asymmetric interaction is
small at $G=G_\infty$, which results in the smooth recovery of the
normal glucose level shown in Fig.~\ref{fig:glucose}.
This result is more evident with the interaction strength $J_1$ larger
and the characteristic time $\tau_G$ shorter. 

If $\alpha$- and $\beta$-cells would inhibit each other, how
should the result change? As suggested already~\cite{Saunders},
the bidirectional inhibitory interactions seem to be optimal in
view of that $\alpha$- and $\beta$-cells play opposite roles in
glucose regulation. Remarkably, however, such symmetric
interactions turn out to result in dynamically unstable responses,
as shown by the dotted line in Fig.~\ref{fig:glucose}. If this
were the case, glucagon secreted by $\alpha$-cells at low glucose
levels would suppress $\beta$-cells from secreting insulin. As the
secretion of insulin decreases, so would the inhibitory effects
of insulin on the glucagon secretion diminish. It should thus
follow that glucagon secretion is not negatively controlled, implying more glucagon secretion. 
Such an apparent positive feedback loop, enhancing hormone secretion, gives rise to
an instability in the islet system (see Fig.~\ref{fig:glucose}).

\subsection{Inhibitory interactions of $\delta$-cells}
\subsubsection{Suppression of co-secretion from $\alpha$- and $\beta$-cells}
There is basal hormone secretion from $\alpha$- and $\beta$-cells
even at the normal glucose level~\cite{Cherrington}, where it is
not necessary to change the blood glucose concentration with the
help of glucagon or insulin. Obviously, the simultaneous secretion
of glucagon and insulin at the normal level should be minimized 
because the opposite effects of the two would cancel out,
nullifying the net effects on the glucose level. Such wasteful
co-secretion of counteracting hormones can be prevented by
$\delta$-cells secreting somatostatin, which inhibits secretion of
both glucagon and insulin.

\begin{figure}
\centerline{
        \begin{tabular}{c}
        \includegraphics*[width=1\textwidth]{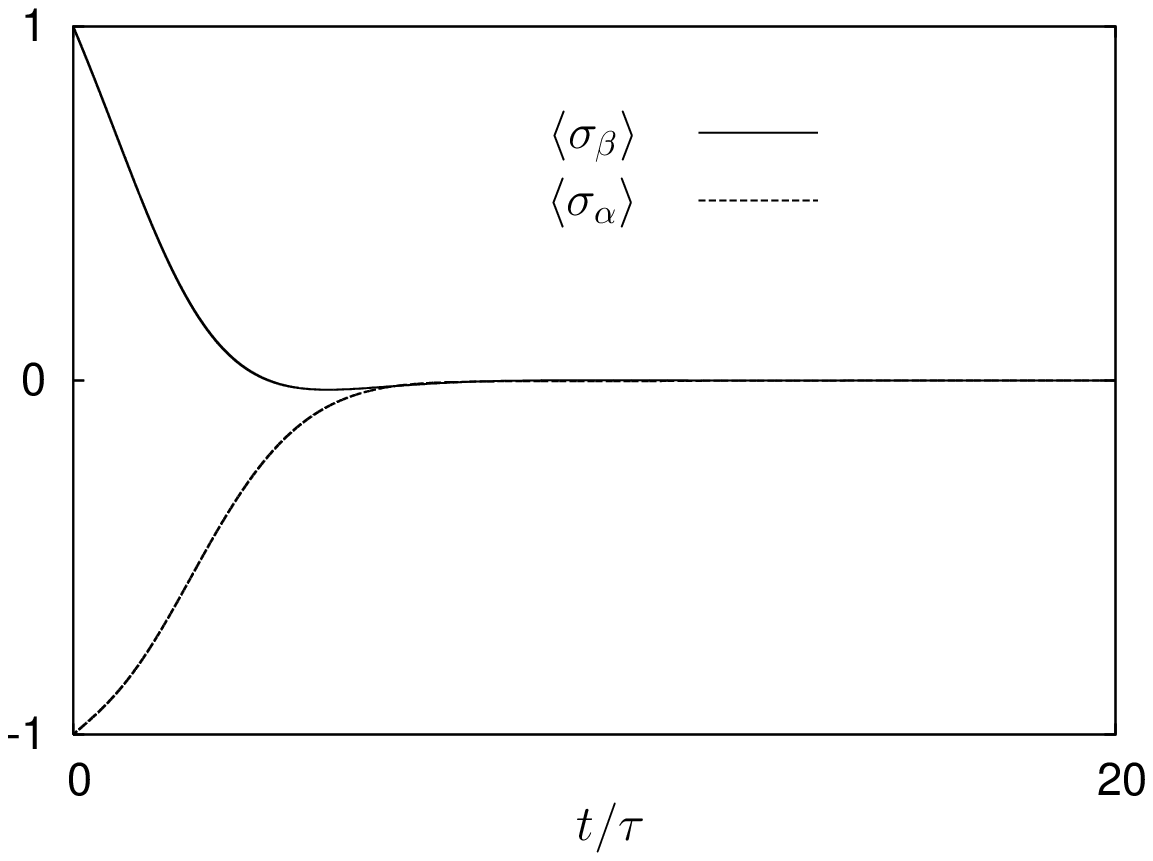} \\
        (a) \\
        \includegraphics*[width=1\textwidth]{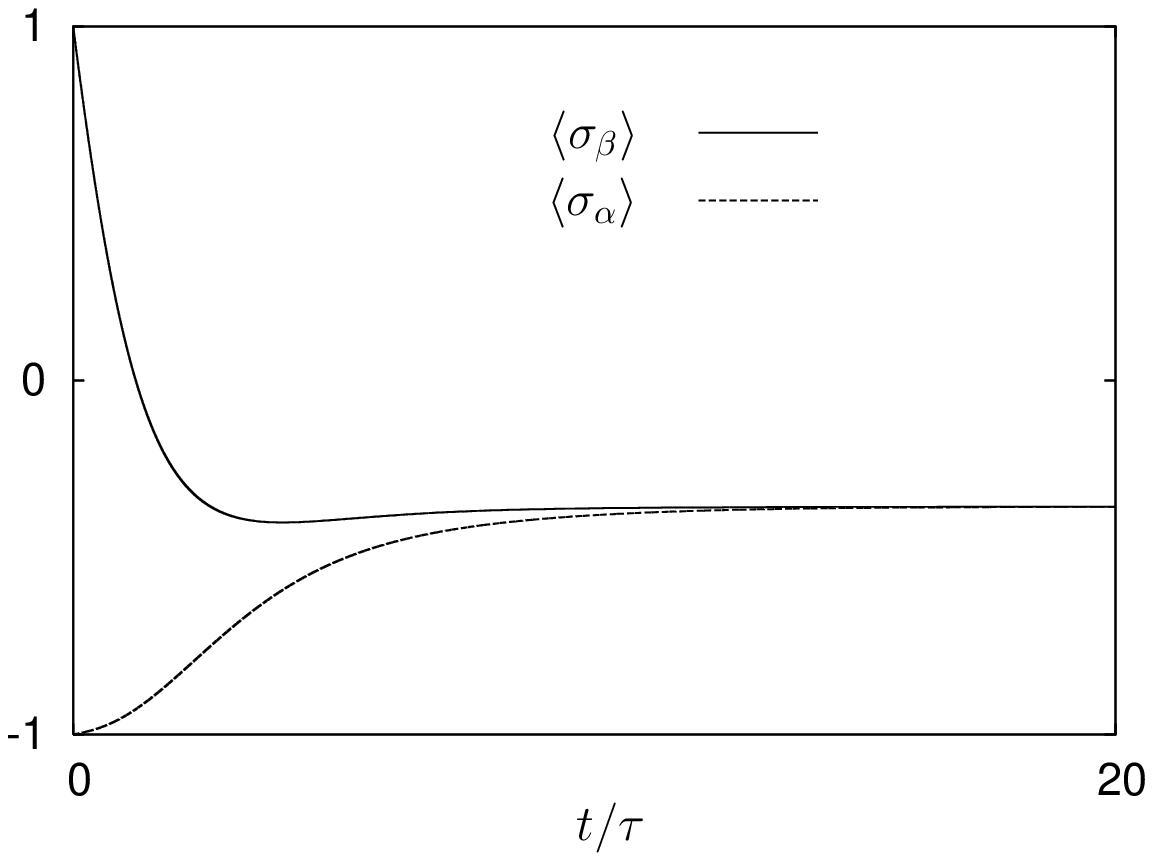} \\
        (b) \\
        \end{tabular}
}
        \caption[]{}
        \label{fig:basal_level}
\end{figure}

In our model, the average activity of cells is given by
$(1+\langle \sigma \rangle)/2$.  Accordingly, the average cell
state $\langle \sigma \rangle = \pm 1$ means that all cells are
active/silent; in particular $\langle \sigma \rangle =0$
corresponds to half of the cells being active. In the absence of the
inhibitory interaction of $\delta$-cells,
Fig.~\ref{fig:basal_level}(a) shows that both $\langle
\sigma_{\alpha} \rangle $ and $\langle \sigma_{\beta} \rangle$
take values greater than $-1$ even at the normal glucose level.
Namely, fluctuations associated with the biological uncertainty
$\Theta$ have some fraction of cells still active, leading to
basal hormone secretion. Here the presence of inhibitory
interactions of $\delta$-cells lowers the basal activity of
$\alpha$- and $\beta$-cells, as shown in
Fig.~\ref{fig:basal_level}(b), which reduces co-secretion of the
counteracting hormones, glucagon and insulin.

\begin{figure}[h]
\centerline{
        \includegraphics*[width=0.8\textwidth]{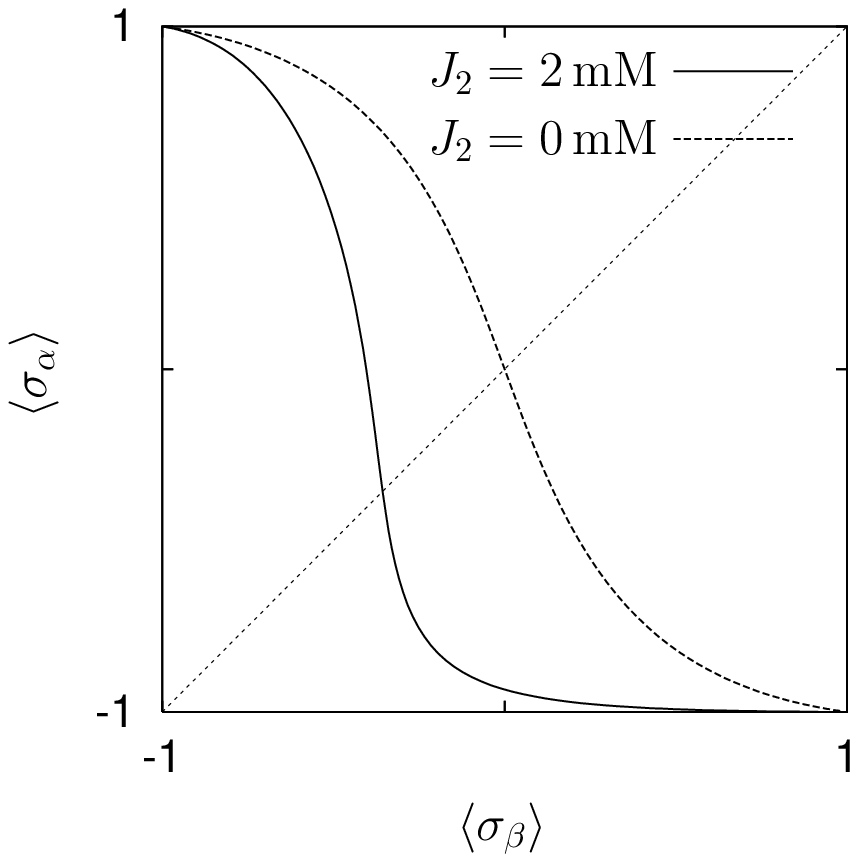}
}
        \caption[]{}
        \label{fig:phasepl}
\end{figure}

Figure~\ref{fig:phasepl} displays the relation between $\langle
\sigma_{\alpha} \rangle$ and $\langle \sigma_{\beta} \rangle$, in
the absence ($J_2 =0$\,mM) and presence ($J_2 =2$\,mM) of the inhibitory
interaction of $\delta$-cells. The system at low or high glucose
levels is described by the upper left or lower right parts of the
curves on the $(\langle \sigma_{\beta} \rangle , \langle
\sigma_{\alpha} \rangle)$ plane, respectively. Namely, when the glucose
concentration is low, $\alpha$- and $\beta$-cells are in high and
in low activity, respectively ($\langle \sigma_{\alpha} \rangle >
0$ and $\langle \sigma_{\beta} \rangle <0$); this is reversed 
at high glucose concentrations.  It is manifested that the inhibitory
interaction of $\delta$-cells reduces simultaneous activation
of $\alpha$- and $\beta$-cells. Compared with the result for
$J_2=0$\,mM (dashed line), the result for $J_2=2$\,mM (solid line) shows
that the activity of $\beta$- or $\alpha$-cells is reduced
substantially at high or low glucose levels.  In particular
$\alpha$-cells remain almost silent ($\langle \sigma_{\alpha}
\rangle \approx -1$) at high glucose levels.
Note, however, that those endocrine cells are not totally silent
at given glucose levels and still exhibit residual activity,
which results from fluctuations in the glucose responses of the
cells. Interestingly, it was suggested that such basal hormone
secretion also plays an effective role: The minimal basal
secretion of glucagon compensates the glucose uptake in the liver
while basal secretion of insulin inhibits over-secretion of the
basal glucagon~\cite{Cherrington}.

\subsubsection{Enhancement of glucose dose-responses of $\beta$-cells}
Another consequence of the inhibitory interaction of
$\delta$-cells is the shift of glucose dose-responses for insulin
secretion to the right direction. This is
associated with the increased control of $\beta$-cells
by $\delta$-cells at high glucose levels. Figure~\ref{fig:dose}
indeed shows that the shift leads to more conspicuous glucose
responses of $\beta$-cells at high glucose levels. In general
$\beta$-cells are coupled with each other through gap-junction
channels, which help the cells synchronize their
behaviors~\cite{Sherman2}. A $\beta$-cell cluster thus tends to
produce all-or-none glucose responses~\cite{Soria}.
In the real islet, on the other hand, $\delta$-cells, with their inhibitory interactions 
depending on the glucose level, can modify the glucose dose-response of $\beta$-cells.
Accordingly, insulin response can be more pronounced at high
glucose levels ($G>0$).

It is observed that some primitive animals have only $\beta$- and
$\delta$-cells in their islets, unlike the mammals whose islets
contain $\alpha$-cells as well as $\beta$- and
$\delta$-cells~\cite{Falkmer}. This difference could perhaps be
attributed to an evolutionary adaptation. At early evolutionary
stages, the islet might be a passive system: Without $\alpha$-cells
directly increasing the glucose level, the glucose level should
increase passively as a result of the decrease in insulin secretion.
Still, the precise glucose dose-responses at high glucose levels
could be possible with $\delta$-cells. At later stages, equipped
with $\alpha$-cells, the islet became an active system with regard
to glucose regulation. It is of interest that this evolutionary
change is correlated with the fact that $\beta$- and $\delta$-cells
are closer to each other than $\alpha$-cells in the development of a
stem cell~\cite{Kemp}. In addition, $\beta$- and $\delta$-cells have
functional similarities of using 
ATP-dependent K$^+$ channels in glucose responses~\cite{Quesada2,Quesada1}.

\section{Discussions}

The islet of Langerhans is a precise system that controls the glucose
level through the use of three main types of endocrine cells.
Here it is of interest to investigate whether the existing interactions
between those cells are beneficial 
for glucose homeostasis.
There are some evidence for the critical role of the interactions,
which may not obviously be addressed by probing $\alpha$- and $\beta$-cells separately.
The molecular mechanism of how $\alpha$-cells regulate glucagon secretion
at variable glucose levels is still not clearly understood~\cite{Gromada}.
Several works attempted to explain this by means of the interactions
between $\alpha$- and $\beta$-cells: At high glucose levels,
glucagon secretion is inhibited by insulin, GABA, or Zn$^{2+}$
secreted from $\beta$-cells~\cite{Gromada,Ishihara}.
There is also a hypothesis that glucose has direct effects on $\alpha$-cells through endoplasmic reticulum Ca$^{2+}$ storage~\cite{Vieira}.
Another evidence for the role of cellular interactions in glucose homeostasis comes from hyperglucagonomia, which occurs in diabetics at abnormally high glucose levels. It appears paradoxical that the glucagon levels of such patients are high even though the blood glucose levels are high enough to make $\alpha$-cells silent~\cite{Gromada}.
This puzzling result can be explained on the basis of cellular interactions
in an islet~\cite{Franklin2,Rorsman,Takahashi}. Note that there is also another explanation of this phenomenon in terms of the peculiar glucose dose-responses of (rat) $\alpha$-cells~\cite{Kemp}.

In contrast, there also exist a few reports that some cellular interactions may not exist and are not necessary for glucose homeostasis: It has been proposed that the microcirculation from $\beta$- to peripheral $\alpha$- and $\delta$-cells prohibits the paracrine action from non-$\beta$ to $\beta$-cells
~\cite{Wayland}.
In addition, it has recently been reported that islet transplantation
is successful in recovering from hyperglycaemia with only $\beta$-cell clusters~\cite{King}.

Nevertheless, the existence of the receptors of signalling molecules such as
insulin, glucagon, somatostatin, glutamate, and GABA, which are expressed in pancreatic endocrine cells, apparently implicates their physiological roles in the fine control of glucose levels~\cite{Gromada,Strowski}.
A better understanding of this tissue, therefore, will contribute to more advanced medical treatment of diabetes than the current one relying mostly
on insulin. For example, it is conceivable to use other hormones such as glucagon and somatostatin for more active and precise glucose control.

A variety of complicated interactions in an islet makes it difficult to
recognize their roles, and existing experiments as to those interactions have
focused mostly on static responses of the endocrine cells.
However, it is likely that the cellular interactions actually contribute
to dynamical responses to glucose.
In this study, therefore, to understand the role of intercellular
interactions between $\alpha$-, $\beta$-, and $\delta$-cells,
we have proposed an islet model and investigated the effects of
integrated intercellular communications between those cells in glucose homeostasis.
Our mathematical model can systematically include all the cellular interactions
and identify their effects on static and/or dynamic responses to external glucose changes.
It also takes individual heterogeneities into consideration,
e.g., in glucose sensitivity; the basal hormone secretion at the normal glucose level
reflects that some cells can be active to secrete hormones even though
most of the cells are silent at that glucose level.
The small variations in glucose responses among homologous cells may contribute crucially to the cellular interactions between 
heterologous 
cells, which are actually activated in quite different glucose concentrations, 
because the heterogeneous responses of homologous cells can lead to an overlap
in the activation between the heterologous cells.
>From this model, it has been revealed that the interactions give more stable, efficient, and accurate control of glucose: First, asymmetric interactions between $\alpha$- and $\beta$-cells contribute to the dynamic stability when the glucose level, perturbed from the normal level, recovers to the latter. Second,
the interactions of somatostatin for glucagon and insulin secretion prevent their wasteful co-secretion
at the normal glucose level. 
In addition, at high glucose levels, the inhibition by $\delta$-cells modifies glucose dose-responses of insulin secretion.
For a more realistic and accurate understanding, 
it would be necessary to know the physiological values of the model parameters. 
In particular, the relative effects of the direct glucose stimulus $G$ 
and paracrine interactions $J$
on the states of endocrine cells should be identified. 

Here it is proposed that these predictions can be verified in experiment.
As for the role of $\delta$-cells, our results may be confirmed
through the use of cell clusters of different compositions of
cell-types, for which the culture method was used in the existing study 
~\cite{Pipeleers2}. Another prediction related with the
asymmetric interactions between $\alpha$- and $\beta$-cells needs to
be verified in vivo experiment on transgenic mice because the
effects should arise in the dynamics of whole-body glucose
regulation. Note that the specific cellular interaction may be blocked
selectively in knockout mice lacking specific hormone receptors in
an endocrine cell~\cite{Diao,Sorensen}.

Beyond the interactions between endocrine cells analyzed in this
study, there exist reports that $\delta$-cells are also influenced
by $\alpha$- and $\beta$-cells~\cite{Unger} and these paracrine
interactions should be considered with the microcirculation of
hormones in an islet as described above~\cite{Wayland}.
It has also been reported that there exist autocrine interactions via which a cell is affected by
its own hormone secretion~\cite{Aspinwall,Cabrera}.
Furthermore, input from 
exocrine cells~\cite{Bertelli,Bishop,Wayland} and glucose-sensing neurons~\cite{Schuit} have been suggested. 
There may thus be more complex communications in the pancreas for glucose homeostasis, 
which are left for further study. Finally, we also point out that the
mathematical model proposed can be generalized to describe cellular
interactions in other systems, e.g., neural networks consisting of
excitatory and inhibitory couplings.

\vspace{1cm}
\noindent
{\bf Acknowledgments} \\
We thank D. Gardner-Hofatt and W. Heuett for useful comments on the manuscript. 
M.Y.C. thanks Asia Pacific Center for Theoretical Physics, where part of this work was performed, for hospitality.
This work was supported in part by the KOSEF/MOST 
grant through National Core Research Center for Systems Bio-Dynamics and
by the KOSEF-CNRS Cooperative Program. 


\clearpage
\section*{Tables}
\begin{table}[h!]
\begin{center}
\caption[Coefficients of transition rates] {Coefficients in the
transition rate. Here $k^x \equiv (g^x + j_1^x + j_2^x + g^x j_1^x
j_2^x)(1+ g^x j_1^x + j_1^x j_2^x + j_2^x g^x)^{-1}$ with $x$
denoting $\alpha$, $\beta$, or $\delta$. Parameters $g^x$, $j_1^x$,
and $j_2^x$ are given in Table~\ref{tab:param}.} \vspace{0.2cm}
\begin{tabular}{cc}
\hline
Coefficient & Value \\
\hline
\vspace{1mm}
$w^x$ & $1 $\\
\vspace{1mm}
$w^x_{\alpha}$ & $k^x$ \\
\vspace{1mm}
$w^x_{\beta}$ & $k^x j^x_1$ \\
\vspace{1mm}
$w^x_{\delta}$ & $k^x j^x_2$ \\
\vspace{1mm}
$w^x_{\alpha \beta}$ & $j^x_1$ \\
\vspace{1mm}
$w^x_{\beta \delta}$ & $ j^x_1 j^x_2$ \\
\vspace{1mm}
$w^x_{\delta \alpha}$ & $j^x_2$ \\
\vspace{1mm}
$w^x_{\alpha \beta \delta}$ & $k^x j^x_1 j^x_2 $ \\
\hline
\end{tabular}
\label{tab:coeff}
\end{center}
\end{table}

\clearpage
\begin{table}[h]
\begin{center}
\caption{Parameters in the coefficients of the transition rate.}
\vspace{0.5cm}
\begin{tabular}{c|ccc}
\hline
$x$ & $\alpha$ & $\beta$ & $\delta$ \\
\hline
$g^{x}$ & ${\rm tanh} (G/2 \Theta)$ & ${\rm tanh} (-G/2 \Theta)$ & ${\rm tanh} (-mG/2 \Theta)$ \\
$j_1^{x}$ & ${\rm tanh} (J_1/4 \Theta)$ & ${\rm tanh} (-J_1/4 \Theta)$ & $0$ \\
$j_2^{x}$ & ${\rm tanh} (J_2/4 \Theta)$ & ${\rm tanh} (J_2/4 \Theta)$ & $0$ \\
\hline
\end{tabular}
\label{tab:param}
\end{center}
\end{table}

\clearpage
\section*{Figure Legends}
\subsubsection*{Figure~\ref{fig:interaction_abd}.}
Schematic diagram of cellular interactions between $\alpha$-,
$\beta$-, and $\delta$-cell{\bf s}. The arrow represents enhancement while
bars represent inhibition.
Here the intercellular interactions between two cells are present only when
both are in active states.

\subsubsection*{Figure~\ref{fig:stationary}.}
Stationary glucose level $G_\infty$, depending on the cellular
interaction strengths $J_1$ and $J_2$.
Note that $G_\infty$ measures the stationary level relative to the
fasting glucose level $\tilde{G}_0$ in the absence of cellular interactions. 

\subsubsection*{Figure~\ref{fig:glucose}.}
Time evolution of the glucose level $G$, depending on the
interactions between $\alpha$- and $\beta$-cells. Starting from the
high-glucose state [$\langle \sigma_\alpha \rangle=-1$ and $\langle
\sigma_\beta \rangle=1$ at $G=1$\,mM (or $\tilde{G} = \tilde{G}_0$ + 1\,mM)], 
the system recovers eventually
the normal glucose level $G_\infty$. The time constants are taken to
be $\tau=\tau_G =1$ for simplicity and the cellular interactions
have strengths $J_1=2$\,mM and $J_2=0$\,mM.

\subsubsection*{Figure~\ref{fig:balance}.}
Balance function for glucose regulation, depending on the
interactions between $\alpha$- and $\beta$-cells. The strengths of
cellular interactions are $J_1=2$\,mM and $J_2=0$\,mM.

\subsubsection*{Figure~\ref{fig:basal_level}.}
Time evolution of the average states $\langle \sigma_\alpha \rangle$
and $\langle \sigma_\beta \rangle$ of $\alpha$- and $\beta$-cells,
starting initially from the high-glucose state $\langle
\sigma_\alpha \rangle = -1$ and $\langle \sigma_\beta \rangle =1$ at
$G=1$\,mM, during the recovery to the normal glucose level. The
asymmetric interactions between $\alpha$- and $\beta$-cells have the
strength $J_1=2$\,mM whereas the inhibitory interactions of
$\delta$-cells are absent in (a) $J_2=0$\,mM but present in (b) $J_2=2$\,mM.

\subsubsection*{Figure~\ref{fig:phasepl}.}
The average state $\langle \sigma_{\alpha} \rangle$ of
$\alpha$-cells versus $\langle \sigma_{\beta} \rangle$ of
$\beta$-cells for the asymmetric interaction $J_1 =2$\,mM, in the
absence ($J_2 =0$\,mM) and presence ($J_2 =2$\,mM) of the inhibitory
interaction of $\delta$-cells.
The dotted line along the diagonal represents the stationary
condition $\langle \sigma_\alpha \rangle = \langle \sigma_\beta
\rangle$.

\begin{figure}[h]
\centerline{
        \includegraphics*[width=1\textwidth]{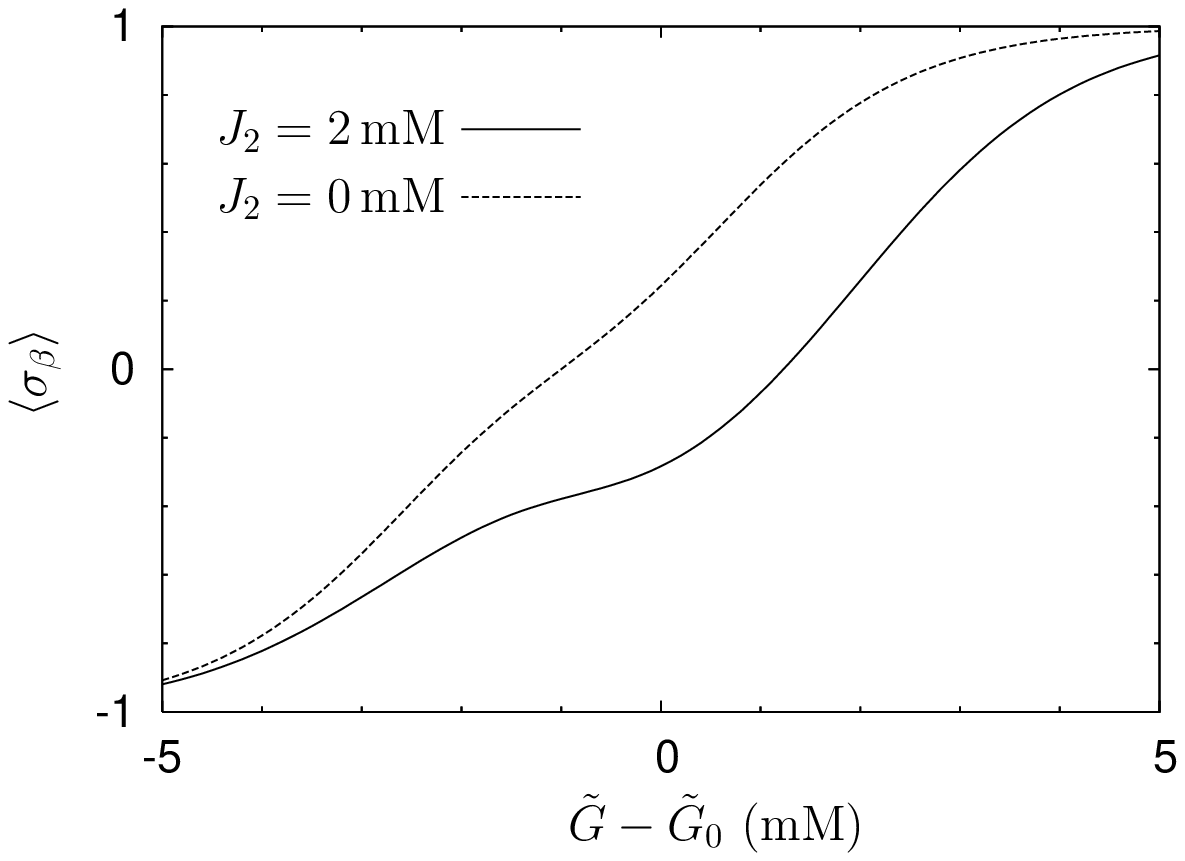}
}
        \caption[]{}
        \label{fig:dose}
\end{figure}

\subsubsection*{Figure~\ref{fig:dose}.}
Glucose dose-responses in the activity of $\beta$-cells for the
inhibitory interaction $J_2 =0$ and $2$\,mM. The asymmetric interactions
are taken to have the strength $J_1=2$\,mM.

\clearpage

\clearpage

\clearpage

\clearpage

\clearpage

\clearpage

\clearpage

\end{document}